\documentclass[a4paper,11pt]{article}
\usepackage[top=1in, bottom=1in, left=1in, right=1in]{geometry}

\usepackage{setspace}
\usepackage{slashbox}

\usepackage[latin1]{inputenc} 
\usepackage{natbib}
\usepackage{amsmath}
\usepackage{amsfonts}
\usepackage{amssymb}
\usepackage{amsthm}
\usepackage{ctable}
\usepackage{subfigure}
\usepackage{graphicx}
\usepackage{cancel}
\usepackage{array}
\usepackage{color}
\usepackage{colortbl} 
\usepackage{multirow} 
\usepackage{subfigure} 
\usepackage{latexsym}
\usepackage{textcomp}
\usepackage{tipa}
\usepackage{url}
\usepackage[hang,small]{caption}

\usepackage{mathtools}
\usepackage{enumitem}

\makeatletter
\renewcommand*\env@matrix[1][*\c@MaxMatrixCols c]{%
  \hskip -\arraycolsep
  \let\@ifnextchar\new@ifnextchar
  \array{#1}}
\makeatother

\def\bzero{\boldsymbol{0}}
\def\b1{\boldsymbol{1}}

\def\bv{\boldsymbol{v}}
\def\bx{\boldsymbol{x}}

\def\bz{\boldsymbol{z}}

\def\bX{\boldsymbol{X}}
\def\bY{\boldsymbol{Y}}
\def\bW{\boldsymbol{W}}

\def\bI{\boldsymbol{I}}

\def\bvartheta{\boldsymbol{\vartheta}}

\def\bmu{\boldsymbol{\mu}}

\def\bpsi{\boldsymbol{\psi}}

\def\balpha{\boldsymbol{\alpha}}

\def\bvartheta{\boldsymbol{\vartheta}}

\def\bpi{\boldsymbol{\pi}}

\def\bSigma{\boldsymbol{\Sigma}}
\def\bGamma{\boldsymbol{\Gamma}}
\def\bLambda{\boldsymbol{\Lambda}}
\def\bOmega{\boldsymbol{\Omega}}
\def\bPsi{\boldsymbol{\Psi}}

\def\diag{\text{diag}}

\begin{document}
\thispagestyle{empty}

\title{Multiple Scaled Contaminated Normal Distribution\\ and Its Application in Clustering} 

\author{Antonio Punzo \and Cristina Tortora}

\date{}

\maketitle

\makeatletter{\renewcommand*{\@makefnmark}{}\footnotetext{
Antonio Punzo, Department of Economics and Business, University of Catania, Catania, Italy (email: \texttt{antonio.punzo@unict.it}). 
Cristina Tortora, Department of Mathematics and Statistics, San Jos\'e State University, California, USA (email: \texttt{cristina.tortora@sjsu.edu}).}\makeatother}


\begin{abstract}
The multivariate contaminated normal (MCN) distribution represents a simple heavy-tailed generalization of the multivariate normal (MN) distribution to model elliptical contoured scatters in the presence of mild outliers, referred to as "bad" points. 
	The MCN can also automatically detect bad points.
	The price of these advantages is two additional parameters, both with specific and useful interpretations: proportion of good observations and degree of contamination. 
	However, points may be bad in some dimensions but good in others. The use of an overall proportion of good observations and of an overall degree of contamination is limiting.
	To overcome this limitation, we propose a multiple scaled contaminated normal (MSCN) distribution with a proportion of good observations and a degree of contamination for each dimension. 
	Once the model is fitted, each observation has a posterior probability of being good with respect to each dimension.   
	Thanks to this probability, we have a method for simultaneous directional robust estimation of the parameters of the MN distribution based on down-weighting and for the automatic directional detection of bad points by means of maximum \textit{a~posteriori} probabilities.
	The term "directional" is added to specify that the method works separately for each dimension. 
	Mixtures of MSCN distributions are also proposed as an application of the proposed model for robust clustering.
	An extension of the EM algorithm is used for parameter estimation based on the maximum likelihood approach.
	Real and simulated data are used to show the usefulness of our mixture with respect to well-established mixtures of symmetric distributions with heavy tails.
\end{abstract}


\section{Introduction}
\label{sec:Introduction}

Statistical inference dealing with continuous multivariate data is commonly focused on the multivariate normal (MN) distribution, with mean $\bmu$ and covariance matrix $\bSigma$, due to its computational and theoretical convenience.
However, for many applied problems, the tails of this distribution are lighter than required.
This is often due to the presence of outliers, i.e., observations that do not comply with the model assumed and that affect the estimation of $\bmu$ and $\bSigma$ \citep{Maro:Yoha:Robu:2016}.
This has created a need for techniques that detect outliers and for which parameter estimates are robust in their presence (see, e.g., \citealp{Devl:Gnan:Kett:Robu:1981} and \citealp{Rous:Lero:Robu:2005}).

Outliers may roughly be divided into two types: mild and gross \citep{Ritt:Robu:2015}.
Outliers are "mild" with respect to the MN distribution (reference distribution) when they do not deviate from the assumed MN model and are not strongly outlying; rather, they produce an overall distribution that is too heavy-tailed to be modeled by the MN. For a discussion of the concept of reference distribution, see \citet{Davi:Gath:Thei:1993}.
Therefore, mild outliers (also referred to as bad points herein, in analogy with \citealp{Aitk:Wils:Mixt:1980}) can be modeled by means of more-flexible distributions, usually symmetric and endowed with heavy tails \citep{Ritt:Robu:2015}.
To define them, the MN distribution is typically embedded in a larger symmetric model with one or more additional parameters denoting the deviation from normality in terms of tail weight.
In this context, the multivariate $t$ (M$t$) distribution (see, e.g., \citealp{Lang:Litt:Tayl:Robu:1989} and \citealp{Kotz:Nada:Mult:2004}), the heavy-tailed versions of the multivariate power exponential (MPE) distribution \citep{Gome:Gome:Main:Nava:Thee:2011}, and the multivariate leptokurtic-normal (MLN) distribution \citep{Bagn:Punz:Zoia:Them:2016}, represent possible symmetric alternatives in the subclass of the elliptically contoured distributions. 

Although the methods discussed above robustify the estimation of $\bmu$ and $\bSigma$ of the reference MN distribution, they do not allow for the automatic detection of bad points.
To overcome this problem, we can consider the multivariate contaminated normal (MCN) distribution of \citet{Tuke:Asur:1960}, a further common and simple elliptically contoured generalization of the MN distribution having heavier tails for the occurrence of bad points;
it is a two-component normal mixture in which one of the components, with a large prior probability $\alpha$, represents the good observations (reference distribution), and the other, with a small prior probability $1-\alpha$, the same mean $\bmu$, and an inflated (with respect to $\eta>1$) covariance matrix $\eta\bSigma$, represents the bad observations \citep{Aitk:Wils:Mixt:1980}.
Advantageously, once the MCN distribution is fitted to the observed data by means of maximum \textit{a~posteriori} probabilities, each observation, if desired \citep{Berk:Bent:Esti:1988}, can be classified as good or bad.
Moreover, bad points are automatically down-weighted in the estimation of $\bmu$ and $\bSigma$.
Thus, the MCN distribution represents a model for the simultaneous robust estimation of $\bmu$ and $\bSigma$ and the detection of mild outliers.

However, the MCN distribution has some drawbacks that are listed below.
\begin{enumerate}[label=(\itshape\alph*\upshape)] 
	\item \label{itm: i} When the scale matrix $\bSigma$ of the MCN distribution is diagonal, the variates are pairwise uncorrelated but can be statistically dependent (with strength of dependence depending on the values of the parameters $\alpha$ and $\eta$).
	\item \label{itm: ii} In relation to the previous point, the product of independent univariate CN distributions, with the same parameters $\alpha$ and $\eta$, is not an MCN distribution.
	\item \label{itm: iii} The MCN distribution, being a normal-scale mixture, belongs to the subclass of elliptically contoured distributions (see, e.g., \citealp{gomez2003survey}, p.~347).
	Thus, its flexibility in terms of shapes is limited. 
	\item \label{itm: iv} Another limitation of the MCN distribution is that all marginals are CN distributions with the same parameters $\alpha$ and $\eta$ and, hence, the same amount of tail weight. 
Therefore, it is not possible to account for very different tail behaviors across dimensions.
	\item \label{itm: v} In terms of robustness, bad points are automatically down-weighted in the maximum likelihood (ML) estimation of $\bmu$ and $\bSigma$ but in the same way for each dimension.
	This does not take into consideration the fact that points may be bad in some dimensions but good in others, a setting that is known in the literature as dimension-wise contamination \citep{Alqa:Prop:2009}. 
	Thus, the down-weighting should be allowed to vary over dimensions.
	\item \label{itm: vi} In relation to the previous point, the procedure to detect outliers induced by the MCN distribution could be defined as \textit{omnibus} in the sense that when a point is detected as bad, it is globally bad. 
	As a practical consequence, once the point is detected as bad, we do not know the dimension(s) yielding this decision.
\end{enumerate}
To overcome these drawbacks, 
we introduce the multiple scaled contaminated normal (MSCN) distribution.
The genesis of our model follows the idea developed by \citet{Forb:Wrai:Anew:2014} to define the multiple scaled $t$ (MS$t$) distribution.
The key elements of the approach are the introduction of a multidimensional Bernoulli variable (indicating whether a point is good or bad separately for each dimension) and the decomposition of $\bSigma$ by eigenvalues and eigenvectors matrices $\bLambda$ and $\bGamma$.  
The result is a distribution in which the scalar parameters $\alpha$ and $\eta$ of the MCN distribution are replaced by two vectors, $\balpha$ and $\boldsymbol{\eta}$, controlling the proportion of good points and the degree of contamination, respectively, separately for each dimension induced by $\bGamma$.

The MSCN distribution offers a remedy to the drawbacks of the MCN distribution discussed above in the following way.
With respect to drawback~\ref{itm: i}, if the scale matrix $\bSigma$ of the MSCN distribution is diagonal, then the variates are independent; as a by-product of this property, the MSCN distribution contains the product of independent univariate CN distributions as a special case, thus providing a remedy to drawback~\ref{itm: ii}. 
With respect to drawback~\ref{itm: iii}, our distribution allows for a greater variety of shapes and, in particular, contours that are symmetric but not necessarily elliptical. 
As concerns drawback~\ref{itm: iv}, the MSCN distribution allows for the parameters $\alpha$ and $\eta$ to be set or estimated differently in each dimension.
It is then possible to account for very different tail behaviors across dimensions.
With respect to drawback~\ref{itm: v}, 
the down-weighting of the observations, in the estimation of $\bmu$ {and $\bLambda$}, is allowed to vary over dimensions (directional robustness). 
Finally, with respect to drawback~\ref{itm: vi}, the procedure to detect outliers induced by the MSCN distribution works separately for each dimension, such that a point may be detected as bad with respect to some dimensions only (directional outlier detection).

The paper is organized as follows. 
Section~\ref{sec:Methodology}, after the recapitulation of some results surrounding the MCN distribution, presents the main contribution of the work (namely, the MSCN distribution and its genesis).
Section~\ref{sec:Mixtures of MSCN Distributions} illustrates the use of the MSCN distribution in robust clustering based on mixture models, which is a further proposal of the present paper. This section also presents a variant of the EM algorithm to fit mixtures of MSCN distributions.
Further computational and operational aspects are discussed in Section~\ref{sec:Further Aspects}. 
Section~\ref{sec:Data Analyses} investigates the performance of the proposed mixture, in comparison with mixtures of some well-established multivariate symmetric distributions with heavy tails, with regard to  artificial and real data. 
Conclusions, as well as avenues for further research, are given in Section~\ref{sec:Discussion and Future Work}. 

\section{Methodology}
\label{sec:Methodology}

\subsection{Preliminaries: The Multivariate Contaminated Normal}
\label{sec:The contaminated Gaussian distribution}

A $d$-variate random vector $\bX=\left(X_1,\ldots,X_d\right)^\top$ is said to follow the multivariate contaminated normal (MCN) distribution with mean vector $\bmu$, scale matrix $\bSigma$, proportion of good points {$\alpha\in\left(0,1\right)$}, and degree of contamination $\eta>1$ if its joint probability density function (pdf) is given by

\begin{equation} 
f_{\text{MCN}}\left(\bx;\bmu,\bSigma,\alpha,\eta\right)=\alpha f_{\text{MN}}\left(\bx;\bmu,\bSigma\right)+\left(1-\alpha\right)f_{\text{MN}}\left(\bx;\boldsymbol{\mu},\eta\boldsymbol{\Sigma}\right),
\label{eq:MCN distribution}
\end{equation}
where $f_{\text{MN}}\left(\cdot;\bmu,\bSigma\right)$ denotes the pdf of a $d$-variate random vector having the multivariate normal (MN) distribution with mean vector $\bmu$ and covariance matrix $\bSigma$.
In the following, when $d=1$, we will substitute the subscripts MN and MCN with N and CN, respectively.
If $\bX$ follows the MCN distribution, we write $\bX \sim \mathcal{CN}_d\left(\bmu,\bSigma,\alpha,\eta\right)$.
As a special case of \eqref{eq:MCN distribution}, if $\alpha$ and $\eta$ tend to one, we obtain the MN distribution with mean vector $\bmu$ and covariance matrix $\bSigma$; in symbols, $\bX\sim \mathcal{N}_d\left(\bmu,\bSigma\right)$.

An advantage of \eqref{eq:MCN distribution} with respect to the multivariate $t$ (M$t$) distribution is that, once the parameters in $\boldsymbol{\vartheta}=\{\bmu,\boldsymbol{\Sigma},\alpha,\eta\}$ are estimated (for example, $\widehat{\boldsymbol{\vartheta}}=\{\widehat{\bmu},\widehat{\bSigma},\widehat{\alpha},\widehat{\eta}\}$), we can establish whether a generic point $\bx^*$ is good via the \textit{a~posteriori} probability
\begin{equation}
P\left(\text{$\bx^*$ is good}\left|\widehat{\boldsymbol{\vartheta}}\right.\right)=
\widehat{\alpha}f_{\text{MN}}\left(\bx^*;\widehat{\bmu},\widehat{\bSigma}\right)\Big/f_{\text{MCN}}\left(\bx^*;\widehat{\bvartheta}\right)
,
\label{eq:probability good}
\end{equation}
and $\bx^*$ will be considered good if $P(\text{$\bx^*$ is good}|\widehat{\boldsymbol{\vartheta}})>1/2$, while it will be considered bad otherwise.

\subsection{Proposal: Multiple Scaled Contaminated Normal}
\label{sec:Multiple Scaled Contaminated Normal Distribution}

In the same spirit of \citet{Forb:Wrai:Anew:2014}, we propose the extension of the MCN distribution to a multiple scaled CN (MSCN) distribution. 
It consists in using the classical eigen decomposition $\bSigma=\bGamma\bLambda\bGamma^\top$ of the scale matrix, where $\bLambda$ is the diagonal matrix of the eigenvalues of $\bSigma$ and $\bGamma$ is a $d\times d$ orthogonal matrix whose columns are the normalized eigenvectors of $\bSigma$, ordered according to their eigenvalues.
Each element in the right-hand side of this decomposition has a different geometric interpretation: $\bLambda$ determines the size and shape of the scatter, while $\boldsymbol{\Gamma}$ determines its orientation.
Moreover, we introduce the indicator variable $V_h$ to be good ($V_h=1$) or bad ($V_h=0$) with respect to the $h$th dimension, $h=1,\ldots,d$, and further define the $d\times d$ diagonal matrix of inverse weights as:
\begin{displaymath}
	\bW_{\bv}=\diag\left\{\left(v_1+\frac{1-v_1}{\eta_1}\right)^{-1},\ldots,\left(v_d+\frac{1-v_d}{\eta_d}\right)^{-1}\right\},
\end{displaymath}
where $\bv=\left(v_1,\ldots,v_d\right)^\top$. 

Based on the quantities introduced above, our MSCN distribution can be written as:
\begin{align}
f_{\text{MSCN}}\left(\bx;\bmu,\bGamma,\bLambda,\balpha,\boldsymbol{\eta}\right)= \sum_{h=1}^d\sum_{v_h=0}^1f_{\text{MN}}\left(\bx;\bmu,\bGamma \bW_{\bv} \bLambda\bGamma^\top\right)p_{v_h}\left(v_h;\alpha_h\right),
\label{eq:MSCN distribution}
\end{align}
where $\balpha=\left(\alpha_1,\ldots,\alpha_d\right)^\top$, $\boldsymbol{\eta}=\left(\eta_1,\ldots,\eta_d\right)^\top$, and
\begin{displaymath}
	p_{\bv}\left(v_1,\ldots,v_d;\balpha\right)=\prod_{h=1}^d p_{v_h}\left(v_h;\alpha_h\right),
\end{displaymath}
with
$p_{v_h}\left(v_h;\alpha_h\right)=\alpha_h^{v_h}\left(1-\alpha_h\right)^{1-v_h}$.
If $\bX$ follows the MSCN distribution, we write $\bX\sim \mathcal{SCN}_d\left(\bmu,\bGamma,\bLambda,\balpha,\boldsymbol{\eta}\right)$. The pdf in \eqref{eq:MSCN distribution} can be equivalently written as: 
\begin{align}  
f_{\text{MSCN}}&\left(\bx;\bmu,\bGamma,\bLambda,\balpha,\boldsymbol{\eta}\right)=\prod_{h=1}^d\left[\alpha_hf_{\text{N}}\left(\left[ \bGamma^\top \left(\bx-\bmu\right)\right]_h;0,\lambda_h\right)\right]^{v_h}\nonumber\\
&\qquad\times\left[\left(1-\alpha_h\right)f_{\text{N}}\left(\left[\bGamma^\top\left(\bx-\bmu\right)\right]_h;0,\eta_h\lambda_h\right)\right]^{\left(1-v_h\right)},
\label{eq:MSCN distribution elegant}
\end{align}
where $\left[\bGamma^\top \left(\bx-\bmu\right)\right]_h$ denotes the $h$th element of the $d$-dimensional vector $\bGamma^\top \left(\bx-\bmu\right)$ and $\lambda_h$ the $h$th diagonal element of $\bLambda$ (or, equivalently, the $h$th eigenvalue of $\bSigma$).

\begin{figure}[!ht]
\centering
\subfigure[$\theta=0$, $\balpha=\left(0.7,0.6\right)^\top$ and $\boldsymbol{\eta}=\left(3,2\right)^\top$\label{fig:Fig1}]
{\includegraphics[width=0.45\textwidth]{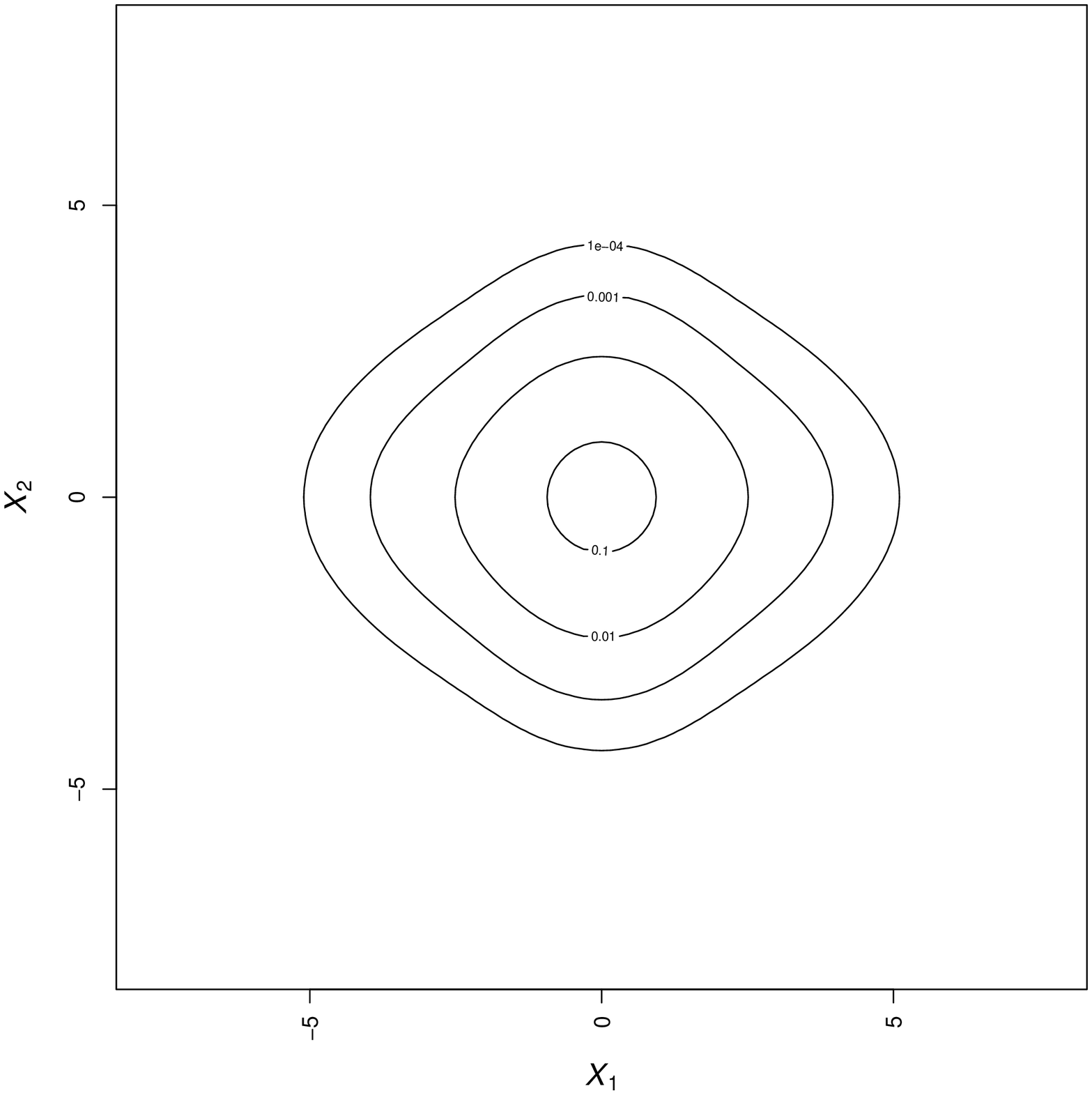}}
\quad
\subfigure[$\theta=\pi/6$, $\balpha=\left(0.7,0.6\right)^\top$ and $\boldsymbol{\eta}=\left(3,2\right)^\top$\label{fig:Fig2}]
{\includegraphics[width=0.45\textwidth]{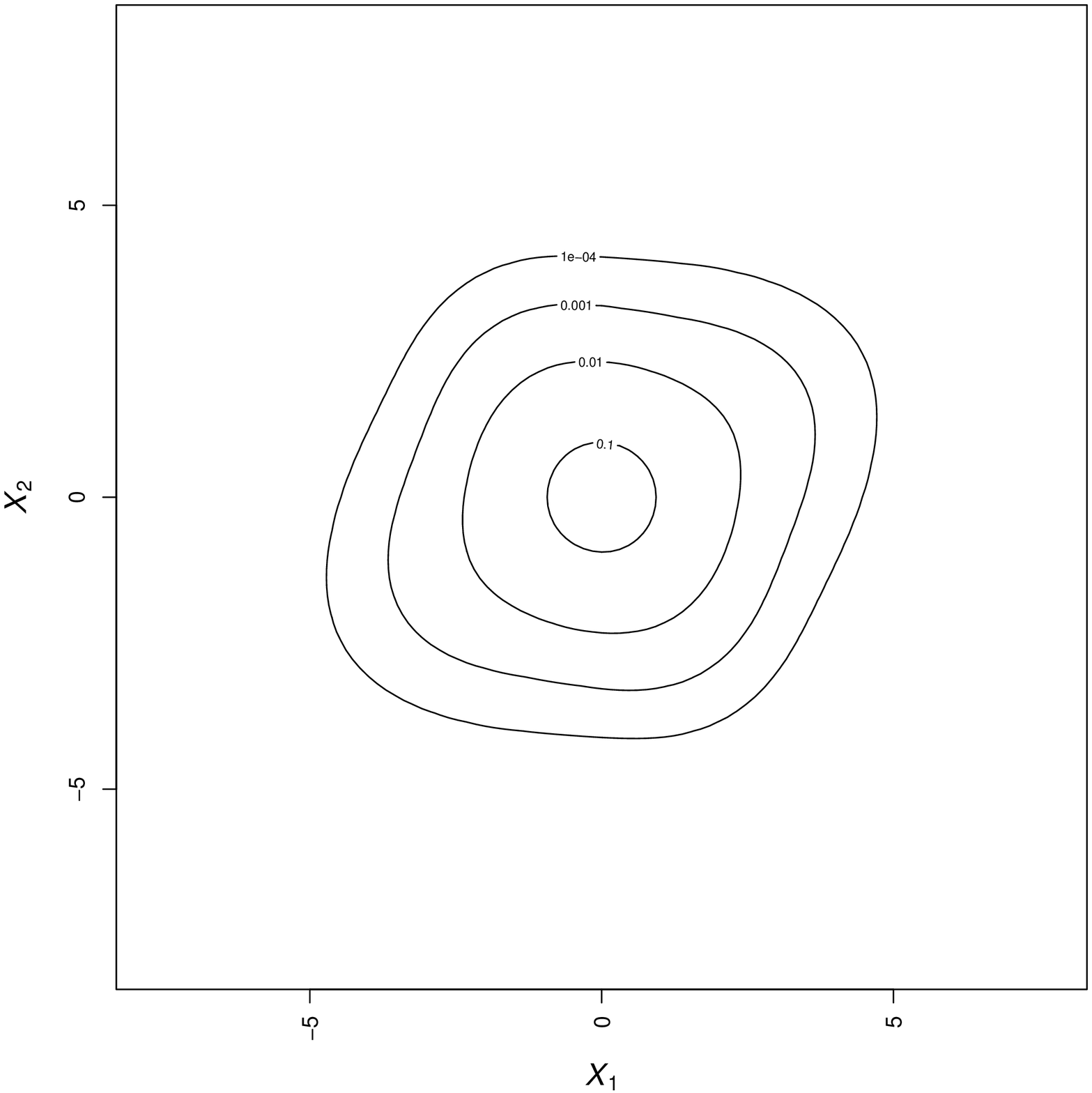}}
\subfigure[$\theta=\pi/6$, $\balpha=\left(0.7,0.7\right)^\top$ and $\boldsymbol{\eta}=\left(10,10\right)^\top$\label{fig:Fig3}]
{\includegraphics[width=0.45\textwidth]{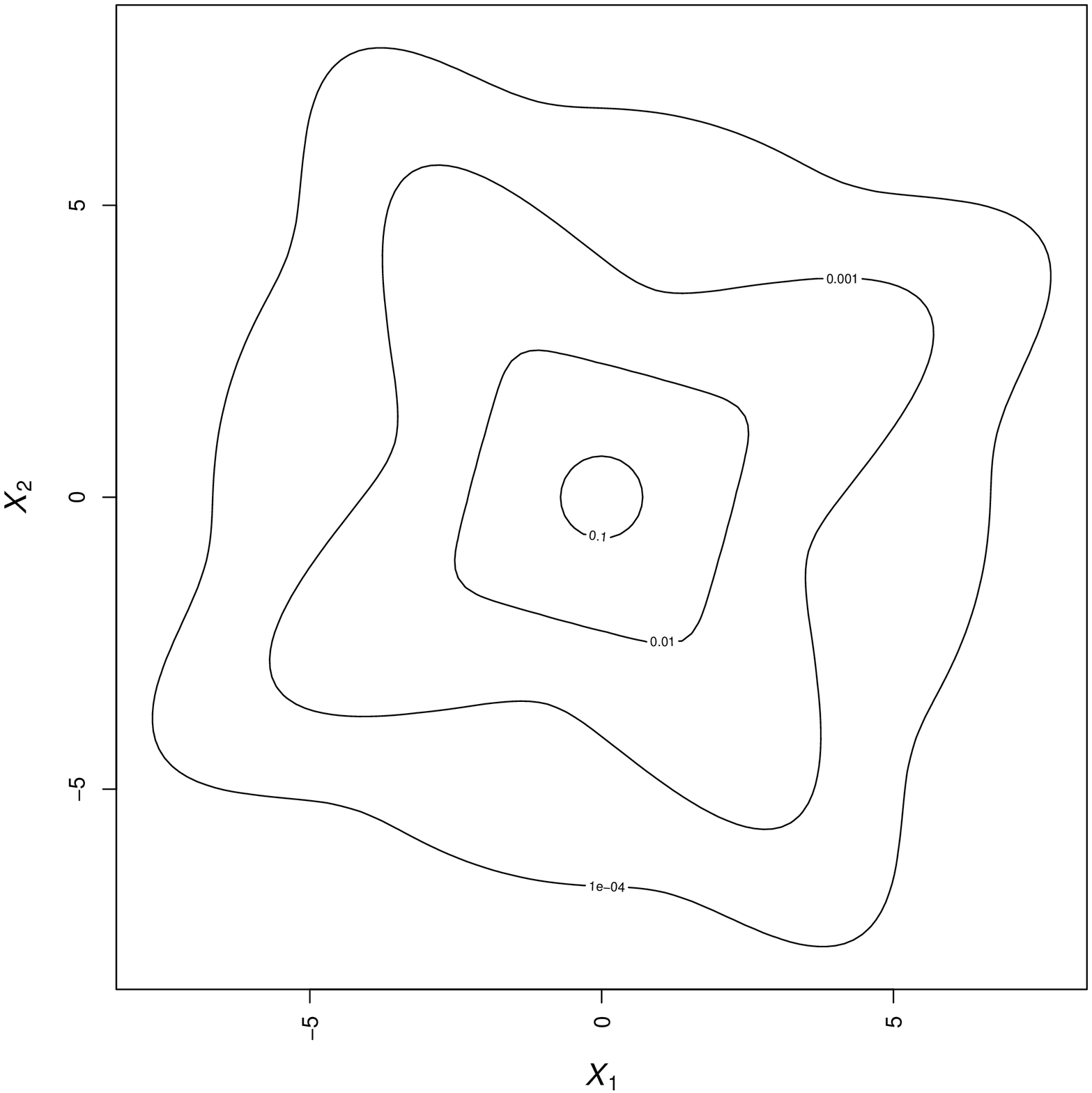}}
\quad
\subfigure[$\theta=\pi/6$, $\balpha=\left(0.95,0.95\right)^\top$ and $\boldsymbol{\eta}=\left(10,10\right)^\top$\label{fig:Fig4}]
{\includegraphics[width=0.45\textwidth]{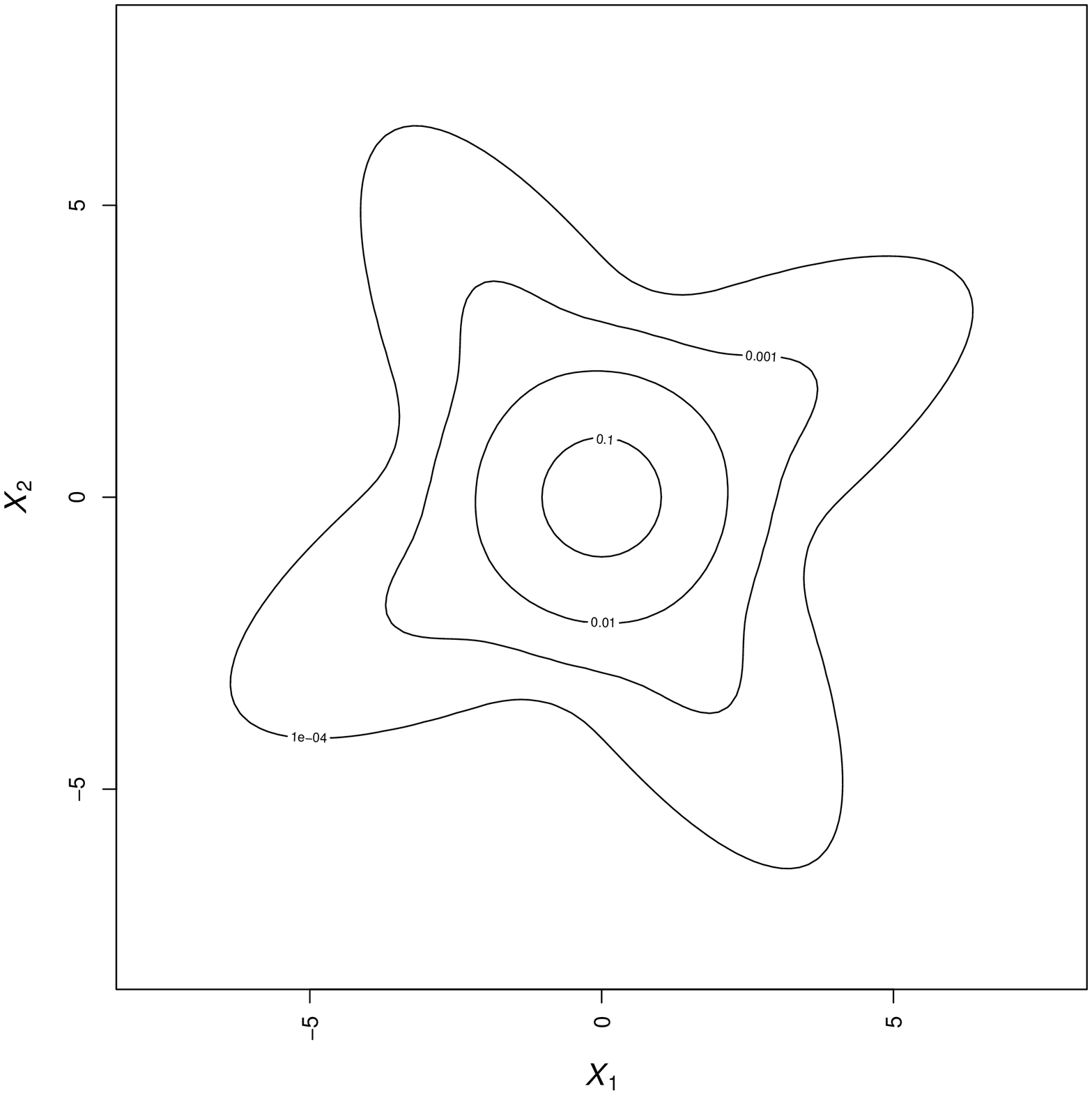}}
\caption{Examples of contour plots of bivariate SCN distributions with $\bmu=\bzero$ and $\bLambda=0.75\bI$.
\label{fig:examples of bivariate SCN distributions}
}
\end{figure}

In the bivariate case ($d=2$), \figurename~\ref{fig:examples of bivariate SCN distributions} shows, via isodensities, some possible shapes of the MSCN distribution by varying $\bGamma$, $\balpha$, and $\boldsymbol{\eta}$, with the mean vector and the eigenvalue matrix fixed, respectively,  to $\bmu=\bzero$ and $\bLambda=0.75\bI$, where $\bI$ denotes the identity matrix.
The orientation matrix $\bGamma$ is seen as a rotation matrix of angle $\theta$, that is 
\begin{displaymath}
\bGamma\left(\theta\right)=
\begin{pmatrix}
\cos\theta & -\sin\theta \\[1ex]
\sin\theta & \cos\theta
\end{pmatrix}.	
\end{displaymath}
\figurename~\ref{fig:examples of bivariate SCN distributions} clearly shows how the shape of the MSCN distribution is not constrained as elliptical, although the symmetry is preserved.
In particular, the choices made for $\balpha$ and $\boldsymbol{\eta}$ produce, among others, "smoothed" rhomboidal (\figurename~\ref{fig:Fig1} and \ref{fig:Fig2}) and starred (\figurename~\ref{fig:Fig3} and \ref{fig:Fig4}) contours.      


Finally, it is easy to show that if $\bY\sim \mathcal{N}_d\left(\bzero,\bI\right)$, then: 
\begin{equation}
\bX=\bmu+\bGamma\bLambda^{\frac{1}{2}}\bW_{\bv}^{\frac{1}{2}}\bY\sim \mathcal{SCN}_d\left(\bmu,\bGamma,\bLambda,\balpha,\boldsymbol{\eta}\right).
\label{eq:alternative MSCN distribution}
\end{equation}
This alternative way to see the MSCN distribution may be useful for random generation.
Moreover, Equation~\eqref{eq:alternative MSCN distribution} makes it easier to see that univariate marginal distributions are linear combinations of CN distributions with the same mean $\mu_h$, $h=1,\ldots,d$, for which, in general, no closed-form expression is available, although it is possible to show that symmetry is preserved.
Therefore, univariate marginal distributions are not in general CN distributions.

\section{The Mixtures of MSCN Distributions}
\label{sec:Mixtures of MSCN Distributions} 

Finite mixtures of distributions are commonly used in statistical modeling as a powerful device for clustering and classification by often assuming that each mixture component represents a cluster (or group) into the original data (see \citealp{McLa:Basf:mixt:1988}, \citealp{Fral:Raft:Howm:1998}, \citealp{Bohn:Comp:2000} and \citealp{mcnicholas16}).

For continuous multivariate random variables, attention is commonly focused on mixtures of MN distributions.
However, in clustering applications, data are often contaminated by mild outliers (see, e.g., \citealp{Bock:Clus:2002}, \citealp{Gall:Ritt:Trim:2009}, and \citealp{Ritt:Robu:2015}), affecting  the estimation of the component means and covariance matrices and the recovery of the underlying clustering structure.
For MN mixtures (MNMs), one of the possible solutions used to deal with mild outliers is the "component-wise" approach: the component MN distributions are separately protected against mild outliers 
by embedding them in more general heavy-tailed, usually symmetric, multivariate distributions.
Examples are M$t$ mixtures (M$t$Ms; \citealp{McLa:Peel:Robu:1998} and \citealp{Peel:McLa:Robu:2000}), MPE mixtures
(MPEMs; \citealp{Zhan:Lian:Robu:2010} and \citealp{Dang:Brow:McNi:Mixt:2015}), MLN mixtures 
\citep{Bagn:Punz:Zoia:Them:2016}, and MS$t$ mixtures 
\citep{Forb:Wrai:Anew:2014}. 
These methods robustify the estimation of the component means and covariance matrices with respect to mixtures of MN distributions, but they do not allow for automatic detection of bad points, although an \textit{a~posteriori} procedure (i.e., a procedure taking place once the model is fitted) to detect bad points with MS$t$Ms is illustrated by \citet{McLa:Peel:fini:2000}.
To overcome this problem, \citet{Punz:McNi:Robu:2016} introduced MCN mixtures (MCNMs); for further recent uses of the MCN distribution in model-based clustering, see \citet{Punz:McNi:Robu:2014,Punz:McNi:RCWM:2017}, \citet{Punz:Maru:Clus:2016}, and \citet{Maru:Punz:Mode:2016}.

\subsection{The Model}
\label{sec:Model}

For a $d$-variate random vector $\bX$, the pdf of a MSCN mixture (MSCNM) with $k$ components can be written as
\begin{equation}
f_{\text{MSCNM}}\left(\bx;\bPsi\right)=\sum_{j=1}^k\pi_jf_{\text{MSCN}}\left(\bx;\bmu_j,\bGamma_j,\bLambda_j,\balpha_j,\boldsymbol{\eta}_j\right),
\label{eq:mixture of MSCN distributions}
\end{equation} 
where $\pi_j$ is the mixing proportion of the $j$th component, with $\pi_j>0$ and $\sum_{j=1}^k\pi_j=1$, $\bmu_j=(\mu_{1j},\ldots,\mu_{dj})^\top$, $\bLambda_j=\diag(\lambda_{1j},\ldots,\lambda_{dj})$, $\balpha_j=(\alpha_{1j},\ldots,\alpha_{dj})^\top$, $\boldsymbol{\eta}_j=(\eta_{1j},\ldots,\eta_{dj})^\top$, 
and $\bPsi$ contains all of the parameters of the mixture.
%
\subsection{Maximum Likelihood Estimation via the AECM Algorithm}
\label{subsec:MSCNM - AECM algorithm}
Let $\bx_1,\ldots,\bx_n$ be a random sample from model~\eqref{eq:mixture of MSCN distributions}.
To find the maximum likelihood (ML) estimates for its parameters $\bPsi$, we adopt the alternating expectation-conditional maximization (AECM) algorithm \citep{Meng:VanD:TheE:1997}. 
It is obtained by combining the expectation-conditional maximization either (ECME) algorithm of \citet{Liu:Rubi:TheE:1994} with the space-alternating generalized EM (SAGE) algorithm of \citet{Fess:Hero:Spac:1994}.
The ECME algorithm is an extension of the classical expectation-maximization (EM) algorithm \citep{Demp:Lair:Rubi:Maxi:1977}, which is a natural approach for ML estimation when the data are incomplete.
The AECM algorithm allows the specification of the complete data to vary where necessary over the conditional maximization (CM) steps, which are a key ingredient of the SAGE algorithm.
As for the ECME algorithm, 
the AECM algorithm monotonically increases the likelihood and reliably converges to a stationary point of the likelihood function (see \citealp{Meng:VanD:TheE:1997} and \citealp{McLa:Kris:TheE:2007}).

In the implementation of the AECM algorithm to fit the MSCNM, we iterate between three steps, one E step and two CM steps, until convergence. 
They arise from the partition $\bPsi=\left\{\bPsi_1,\bPsi_2\right\}$, where $\bPsi_1$ contains $\pi_j$, $\balpha_j$, $\bmu_j$ and $\boldsymbol{\eta}_j$, while $\bPsi_2$ contains $\bGamma_j$ and $\bLambda_j$, $j=1,\ldots,k$.

In the first CM-step, where $\bPsi_1$ is updated, we have a two-level source of incompleteness.
The first-level source of incompleteness, the classical one in the use of mixture models, arises from the fact that for each observation, we do not know its component membership; this source is governed by an indicator vector $\bz_i=\left(z_{i1},\ldots,z_{ik}\right)$, where $z_{ij}=1$ if $\bx_i$ comes from component $j$ and $z_{ij}=0$ otherwise.
The second-level source of incompleteness arises from the fact that we do not know if the generic observation $\bx_i$, $i=1,\ldots,n$ is good or bad with respect to the generic dimension $h$, $h=1,\ldots,d$ and to the $j$th group, $j=1,\ldots,k$; this source of incompleteness is governed by a $n \times d \times k$ indicator array with elements $v_{ihj}$, $i=1,\ldots,n$, $h=1,\ldots,d$, and $j=1,\ldots,k$, where $v_{ihj}=1$ if $x_{ih}$ is good with respect to the $j$th component and $v_{ihj}=0$ otherwise.
The values of $z_{ij}$ and $v_{ihj}$ are used for the definition of the following two-level complete-data likelihood 
\begin{align}
&L_{c1}\left(\bPsi\right)=\prod_{i=1}^n\prod_{j=1}^k
\pi_j\prod_{h=1}^d\left[\alpha_{hj}f_{\text{N}}\left(\left[\bGamma_j^\top \left(\bx_i-\bmu_j\right)\right]_h;0,\lambda_{hj}\right)\right]^{v_{ihj}}\nonumber\\
&\qquad \times \left[\left(1-\alpha_{hj}\right)f_{\text{N}}\left(\left[\bGamma_j^\top \left(\bx_i-\bmu_j\right)\right]_h;0,\eta_{hj}\lambda_{hj}\right)\right]^{\left(1-v_{ihj}\right)}.
\label{eq:mixture - complete-data likelihood}
\end{align}
The corresponding two-level complete-data log-likelihood can be so written as
\begin{equation}
l_{c1}\left(\bPsi\right)=l_{c1,1}\left(\bpi\right)+\sum_{j=1}^k\left[l_{c1,2j}\left(\balpha_j\right)+l_{c1,3j}\left(\bpsi_j\right)\right],
\label{eq:mixture - complete-data log-likelihood}
\end{equation}
where $\bpi=\left(\pi_1,\ldots,\pi_k\right)^\top$ and 
\begin{equation}
l_{c1,1}\left(\bpi\right) = \sum_{i=1}^{n}\sum_{j=1}^k{z}_{ij}\ln \pi_j, \label{eq:mixture MSCN complete loglik pi}
\end{equation}
\begin{equation}
l_{c1,2j}\left(\balpha_j\right) = \sum_{i=1}^{n}\sum_{h=1}^dz_{ij}\left[v_{ihj}\ln \alpha_{hj}+\left(1-v_{ihj}\right)\ln \left(1-\alpha_{hj}\right)\right],
\label{eq:mixture MSCN complete loglik alpha}\\ 
\end{equation}
\begin{equation}
l_{c1,3j}\left(\bpsi_j\right) =
-\frac{1}{2}\sum_{i=1}^n\sum_{h=1}^dz_{ij}
\left\{
\ln \lambda_{hj} +\left(1-v_{ihj}\right)\ln\eta_{hj} 
+\left(v_{ihj}+\frac{1-v_{ihj}}{\eta_{hj}}\right)\frac{\left[\bGamma_j^\top \left(\bx_i-\bmu_j\right)\right]_h^2}{\lambda_{hj}}\right\}, 
\label{eq:mixture MSCN complete loglik other parameters}
\end{equation}
with $\bpsi_j=\left\{\bmu_j,\bGamma_j,\bLambda_j,\boldsymbol{\eta}_j\right\}$, $j=1,\ldots,k$. 

The second CM step, in which we update $\bPsi_2$, is based on a single-level source of incompleteness that refers to the indicator vector $\bz_i$, $i=1,\ldots,n$.
The single-level complete-data likelihood of this step is:
\begin{equation*}
L_{c2}\left(\bPsi\right)=\prod_{i=1}^n\prod_{j=1}^k\pi_jf_{\text{MSCN}}\left(\bx_i;\bmu_j,\bGamma_j,\bLambda_j,\balpha_j,\boldsymbol{\eta}_j\right),
\end{equation*}  
and the corresponding single-level complete-data log-likelihood is:
\begin{equation*}
l_{c2}\left(\bPsi\right) = l_{c2,1}\left(\bpi\right) + \sum_{j=1}^k l_{c2,2j}\left(\bmu_j,\bGamma_j,\bLambda_j,\balpha_j,\boldsymbol{\eta}_j\right),
\end{equation*} 
where $l_{c2,1}\left(\bpi\right)=l_{c1,1}\left(\bpi\right)$ and: 
\begin{equation}
l_{c2,2j}\left(\bmu_j,\bGamma_j,\bLambda_j,\balpha_j,\boldsymbol{\eta}_j\right)=\sum_{i=1}^nz_{ij}\ln f_{\text{MSCN}}\left(\bx_i;\bmu_j,\bGamma_j,\bLambda_j,\balpha_j,\boldsymbol{\eta}_j\right).
\label{eq:CM-step 2 - one-level likelihood - component j}
\end{equation}

The three steps of our AECM algorithm, for the generic $\left(r+1\right)$th iteration, $r=0,1,\ldots$, are detailed below.
How it will be better understood after the reading of the following steps, when $k=1$ we obtain an ECME algorithm as a special case of our AECM algorithm.


\subsubsection{E-step}

The E-step requires the calculation of:
\begin{equation*}
E_{\bPsi^{\left(r\right)}}\left(Z_{ij}\left|\bx_i\right.\right)
=\frac{\pi_j^{\left(r\right)}f_{\text{MSCN}}\left(\bx_i;\bmu_j^{\left(r\right)},\bGamma_j^{\left(r\right)},\bLambda_j^{\left(r\right)},\balpha_j^{\left(r\right)},\boldsymbol{\eta}_j^{\left(r\right)}\right)}{f_{\text{MSCNM}}\left(\bx_i;\bPsi^{\left(r\right)}\right)}\eqqcolon z_{ij}^{\left(r\right)},
\end{equation*}
which is the posterior probability that $\bx_i$ belongs to the $j$th component of the mixture using the current fit $\bPsi^{\left(r\right)}$ for $\bPsi$ and:
\begin{equation}
E_{\bvartheta_j^{\left(r\right)}}\left(V_{ihj}|\bx_i,Z_{ij}=1\right)=
\frac{
\alpha_{hj}^{\left(r\right)}f_{\text{N}}\left(\left[{\bGamma_j^{\left(r\right)}}^\top \left(\bx_i-\bmu_j^{\left(r\right)}\right)\right]_h;0,\lambda_{hj}^{\left(r\right)}\right)
}{
f_{\text{CN}}\left(\left[{\bGamma_j^{\left(r\right)}}^\top\left(\bx_i-\bmu_j^{\left(r\right)}\right)\right]_h;0,\lambda_{hj}^{\left(r\right)},\alpha_{hj}^{\left(r\right)},\eta_{hj}^{\left(r\right)}\right)
}\eqqcolon v_{ihj}^{\left(r\right)},	
\label{eq:AECM - v}
\end{equation}
which is the posterior probability that $\bx_i$ is good with respect to the $h$th dimension in the $j$th mixture component using the current fit $\bvartheta_j^{\left(r\right)}$ for $\bvartheta_j=\left\{\bmu_j,\bGamma_j,\bLambda_j,\balpha_j,\boldsymbol{\eta}_j\right\}$, $i=1,\ldots,n$, $h=1,\ldots,d$, and $j=1,\ldots,k$.
Then, by substituting $z_{ij}$ with $z_{ij}^{\left(r\right)}$ and $v_{ihj}$ with $v_{ihj}^{\left(r\right)}$ in \eqref{eq:mixture MSCN complete loglik pi}--\eqref{eq:mixture MSCN complete loglik other parameters}, we obtain the functions to be maximized in the CM-steps, at the $\left(r+1\right)$th iteration, to obtain the updates for the parameters of the model.


\subsubsection{CM Step 1}
\label{subsec:MSCNM CM Step 1}

At the first CM step, the maximization of the expected counterpart of $l_{c1}$ in \eqref{eq:mixture - complete-data log-likelihood} with respect to $\pi_j$, $\balpha_j$, $\bmu_j$ and $\boldsymbol{\eta}_j$, $j=1,\ldots,k$, with $\bGamma_j$ and $\bLambda_j$ fixed at $\bGamma_j^{\left(r\right)}$ and $\bLambda_j^{\left(r\right)}$, respectively, yields 
\begin{align}
\pi_j^{\left(r+1\right)}=
&
\displaystyle\frac{n_j^{\left(r\right)}}{n},\nonumber\\ 
\alpha_{hj}^{\left(r+1\right)}=&
\frac{1}{n_j^{\left(r\right)}}\sum_{i=1}^n z_{ij}^{\left(r\right)}v_{ihj}^{\left(r\right)},
\label{eq:AECM alpha}
\\
\mu_{hj}^{\left(r+1\right)} =
&
\displaystyle\sum_{i=1}^n
\frac{
z_{ij}^{\left(r\right)}\left(v_{ihj}^{\left(r\right)}+\displaystyle\frac{1-v_{ihj}^{\left(r\right)}}{\eta_{hj}^{\left(r\right)}}\right) 
}{
\displaystyle\sum_{l=1}^n z_{lj}^{\left(r\right)}\left(v_{lhj}^{\left(r\right)}+\frac{1-v_{lhj}^{\left(r\right)}}{\eta_{hj}^{\left(r\right)}}\right)
} 
x_{ih}, \label{eq:AECM mu}\\
\eta_{hj}^{\left(r+1\right)}=&\max\left\{\eta^*,\displaystyle\sum_{i=1}^n \frac{z_{ij}^{\left(r\right)}\left(1-v_{ihj}^{\left(r\right)}\right)}{\displaystyle\sum_{l=1}^n z_{lj}^{\left(r\right)}\left(1-v_{lhj}^{\left(r\right)}\right)} \frac{\left[\left(\bGamma_j^{\left(r\right)}\right)^\top\left(\bx_i-\bmu_j^{\left(r+1\right)}\right)\right]_h^2}{\lambda_{hj}^{\left(r\right)}}\right\}, \nonumber
\end{align}
where $n_j^{\left(r\right)}=\displaystyle\sum_{i=1}^n z_{ij}^{\left(r\right)}$ ad $\eta^*$ is a number close to 1 from the right, $j=1,\ldots,k$ and $h=1,\ldots,d$; for the analyses herein, we use $\eta^*=1.001$.

\subsubsection{CM Step 2}
\label{subsec:MSCNM CM Step 2}

The updates of $\bGamma_j$ and $\bLambda_j$, $j=1,\ldots,k$, are obtained at the second CM step by maximizing 
\begin{equation}
Q_{2,2j}\left(\bGamma_j,\bLambda_j\right)=\sum_{i=1}^nz_{ij}^{\left(r\right)}\ln f_{\text{MSCN}}\left(\bx_i;\bmu_j^{\left(r\right)},\bGamma_j,\bLambda_j,\balpha_j^{\left(r\right)},\boldsymbol{\eta}_j^{\left(r\right)}\right),
\label{eq:AECM - CM-step}
\end{equation}
the expected counterpart of $l_{c2,2j}$ in \eqref{eq:CM-step 2 - one-level likelihood - component j}, with $\balpha_j$, $\bmu_j$ and $\boldsymbol{\eta}_j$ fixed at $\balpha_j^{\left(r+1\right)}$, $\bmu_j^{\left(r+1\right)}$ and $\boldsymbol{\eta}_j^{\left(r+1\right)}$, respectively. 
The function in \eqref{eq:AECM - CM-step} is equivalent to the observed-data log-likelihood function for the MSCN distribution, with the exception that each observation $\bx_i$ contributes to the log likelihood with a known weight $z_{ij}^{\left(r\right)}$.
To obtain the updates $\bGamma_j^{\left(r+1\right)}$ and $\bLambda_j^{\left(r+1\right)}$, $j=1,\ldots,k$, 
$Q_{2,2j}$ is maximized with respect to a transformation of $\bGamma_j$ and $\bLambda_j$, i.e., $\bSigma_j=\bGamma_j\bLambda_j\bGamma_j^\top$; the Cholesky decomposition $\bSigma_j=\bOmega^\top\bOmega$ is considered to make the maximization unconstrained, and the updates $\bGamma_j^{\left(r+1\right)}$ and $\bLambda_j^{\left(r+1\right)}$ are finally obtained by back-transformation.
Operationally, this is done via the \texttt{optim()} function for \textsf{R}. 
The BFGS method or algorithm, passed to \texttt{optim()} via the argument \texttt{method}, is used for maximization.

\section{Further Computational and Operational Aspects}
\label{sec:Further Aspects}

\subsection{Initialization}
\label{subsec:Initialization}

As is well-documented in literature, the starting values impact the results of any variant of the EM algorithm; therefore, their choice constitutes a very important issue (see, e.g., \citealp{Bier:Cele:Gova:Choo:2003}, \citealp{Karl:Xeka:Choo:2003}, and \citealp{Bagn:Punz:Fine:2013}).

We decided to start our AECM algorithm by the first CM step.
This implies the need of initial quantities $z_{ij}^{(0)}$ and $v_{ihj}^{(0)}$ for the E step and $\bLambda_j^{(0)}$ and $\bGamma_j^{(0)}$ for the second CM step.
For $z_{ij}^{(0)}$, we tried several options: $k$-means, $k$-medoids, and multivariate normal mixture.
The best one, used in the data analyses of Section~\ref{sec:Data Analyses}, was the partition arising from a preliminary run of the $k$-medoids method \citep{KauRou90b}.
Finally, we fix $v_{ihj}^{(0)}=0.99$ and define $\bLambda_j^{(0)}$ and $\bGamma_j^{(0)}$ as the eigenvalues and eigenvectors matrices, respectively, of the $j$th cluster covariance matrix.

\subsection{Convergence Criterion}
\label{subsec:Convergence Criterion}

A stopping criterion based on Aitken's acceleration \citep{Aitk:OnBe:1926} is used to determine convergence of the algorithms illustrated in Section~\ref{subsec:MSCNM - AECM algorithm}. 
The commonly used stopping rules can yield convergence earlier than the Aitken stopping criterion, resulting in estimates that might not be close to the ML estimates. 
The Aitken acceleration at iteration $r$ is:
\begin{equation*}
a^{\left(r\right)}=\frac{l^{\text{new}}-l^{\left(r\right)}}{l^{\left(r\right)}-l^{\left(r-1\right)}},
\end{equation*}
where $l^{\left(r\right)}$ is the (observed-data) log likelihood value from iteration $r$. 
An asymptotic (with respect to the iteration number) estimate of the log likelihood at iteration $r + 1$ can be computed via:
\begin{equation*}	
l_A^{\text{new}}=l^{\left(r\right)}+\frac{1}{1-a^{\left(r\right)}}\left(l^{\text{new}}-l^{\left(r\right)}\right);
\end{equation*}
cf.~\citet{Bohn:Diet:Scha:Schl:Lind:TheD:1994}. 
Convergence is assumed to have been reached when $l_A^{\text{new}}-l^{\left(r\right)}<\epsilon$, provided that this difference is positive (cf.~\citealp{Lind:Mixt:1995}, \citealp{McNi:Murp:McDa:Fros:Seri:2010}, and \citealp{Sube:Punz:Ingr:McNi:Clus:2013,Sube:Punz:Ingr:McNi:Clus:2015}). 
We use $\epsilon = 0.001$ in the analyses herein and set the maximum number of iterations to 200.

\subsection{Some Notes on Directional Robustness}
\label{subsec:Some Notes on Robustness}

The MSCN mixture model provides improved directional estimates (robust directional estimates) of the $d$ dimensions of $\bmu_j$, $j=1,\ldots,k$, in the presence of mild outliers.
This is made possible because the influence of $x_{ih}$, the $h$th dimension of $\bx_i$ assigned to the $j$th cluster, is reduced (down-weighted) as the squared Mahalanobis distance
$$
\delta_{ihj}=\left[\bGamma_j^\top \left(\bx_i-\bmu_j\right)\right]_h^2\Big/\lambda_{hj}
$$
increases.
This is the underlying idea of $M$ estimation \citep{Maro:Robu:1976}, which uses a decreasing weighting function $w\left(\delta_{ihj}\right) : \left(0,\infty\right) \rightarrow \left(0,\infty\right)$ to down-weight the observations $x_{ih}$ with large $\delta_{ihj}$ values.
	To be more precise, according to \eqref{eq:AECM mu}, $\mu_{hj}^{(r+1)}$ can be viewed because $\alpha_{hj}$ and $\eta_{hj}$ are estimated from the data by ML, as an adaptively weighted sample mean, in the sense used by \citet{Hogg:Adap:1974}, with weights of:
	\begin{equation}
v_{ihj}^{(r)}+\displaystyle\frac{1-v_{ihj}^{(r)}}{\eta_{hj}^{(r)}}.
\label{eq:weights}
\end{equation}
This approach, in addition to be a type of $M$ estimation, in each cluster and each dimension follows \citet{Box:Samp:1980} and \citet{Box:Tiao:Baye:2011} in embedding the normal model in a larger model with one or more parameters (here $\alpha_{hj}$ and $\eta_{hj}$) that afford protection against non-normality.
For a discussion on down-weighting for the contaminated normal distribution, see also \citet{Litt:Robu:1988} and \citet{Punz:McNi:Robu:2016}.

Below, we make explicit the formulation of our weighting function and demonstrate its decreasing behavior with respect to $\delta_{ihj}$.
If we substitute $v_{ihj}^{(r)}$ in \eqref{eq:weights} with its explicit formulation given in \eqref{eq:AECM - v}, avoid the use of the iteration superscript and use the simplified notation $\delta_{ihj}$ for the squared Mahalanobis distance and then the weighting function of our approach results:
\begin{equation}
w\left(\delta_{ihj};\alpha_{hj},\eta_{hj}\right)=
1 + \frac{\left(1-\alpha_{hj}\right)\left(\eta_{hj} - 1\right)e^{\frac{\delta_{ihj}}{2}}}{(\alpha_{hj}-1) \eta_{hj} e^{\frac{\delta_{ihj}}{2}} - \alpha_{hj} \sqrt{\eta_{hj}^3} e^{\frac{\delta_{ihj}}{2 \eta_{hj}}}}.
\label{eq:downweight function}
\end{equation}
The first order derivative of $w\left(\delta_{ihj};\alpha_{hj},\eta_{hj}\right)$ is:
\begin{equation}
w^{\prime}\left(\delta_{ihj};\alpha_{hj},\eta_{hj}\right) = -
\frac{\alpha_{hj} \left(1-\alpha_{hj}\right) \left(\eta_{hj} - 1\right)^2 e^{\frac{\delta_{ihj} \left(\eta_{hj} + 1\right)}{2 \eta_{hj}}}}{2 \sqrt{\eta_{hj}^3} \left[\left(\alpha_{hj} - 1\right) e^{\frac{\delta_{ihj}}{2}}-\alpha_{hj} \sqrt{\eta_{hj}} e^{\frac{\delta_{ihj}}{2\eta_{hj}}}\right]^2}.
\label{eq:partial derivatives - delta}
\end{equation}
Due to the constraints $\alpha_{hj}\in\left(0,1\right)$ and $\eta_{hj}>0$, it is straightforward to realize that $w^{\prime}\left(\delta_{ihj};\alpha_{hj},\eta_{hj}\right)$ is always negative, and this implies that $w\left(\delta_{ihj};\alpha_{hj},\eta_{hj}\right)$ is a decreasing function of $\delta_{ihj}$.
For further details about down-weighting with mixture models based on the contaminated normal distribution, see \citet{Punz:McNi:Robu:2016} and \citet{Mazz:Punz:Mixt:2018}. 

\subsection{Constraints for Directional Detection of Bad Points}
\label{subsec:Constraints for Detection of Outliers}

When our MSCN mixture is used for the directional detection of bad points, $\left(1-\alpha_{hj}\right)$ and $\alpha_{hj}$ represent the proportion of bad points and the degree of contamination, respectively, in the $h$th dimension and $j$th group.
Then, for the former parameter, one could require that the proportion of good data is at least equal to a predetermined value, $\alpha^*$.
In this case, it is easy to show \citep{JSS:ContaminatedMixt} that the update for $\alpha_{hj}$ in \eqref{eq:AECM alpha} becomes:
\begin{equation*}
\alpha_{hj}^{\left(r+1\right)}=\max\left\{\alpha^*,\frac{1}{n_j^{(r)}}\sum_{i=1}^n z_{ij}^{\left(r\right)}v_{ihj}^{\left(r\right)}\right\}.
\end{equation*}
In the data analyses of Section~\ref{sec:Data Analyses}, we use this approach to update $\alpha_{hj}$, and we take $\alpha^*=0.5$.
The value 0.5 is justified because, in robust statistics, it is usually assumed that at least half of the points are good \citep{Punz:McNi:Robu:2016}.

\subsection{Automatic directional detection of outliers}
\label{subsec:MSCNMs Directional outlier detection}

Here, we illustrate how the automatic directional detection of bad points works for the more general MSCN mixture model introduced in Section~\ref{sec:Model}.
In detail, the classification of a generic observation $\bx_i$, according to model~\eqref{eq:mixture of MSCN distributions}, means: 
\begin{description}
	\item[step 1.] determine its cluster of membership; and
	\item[step 2.] establish whether its generic $h$th dimension $x_{ih}$, $h=1,\ldots,d$, is good or bad in that cluster.
\end{description}
Let $\widehat{z}_{ij}$ and $\widehat{v}_{ihj}$ be the values of $z_{ij}^{\left(r\right)}$ and $v_{ihj}^{\left(r\right)}$, respectively, at convergence of the AECM algorithm.
To evaluate the cluster membership of $\bx_i$, we use, as is typical in model-based clustering applications, the maximum \textit{a~posteriori} (MAP) classification, i.e.,
$$
\text{MAP}(\widehat{z}_{ij})=
\begin{cases}
1 & \text{if } \max_g\{\widehat{z}_{ig}\} \text{ occurs in cluster $j$,}\\
0 & \text{otherwise.}
\end{cases}
$$
We then consider $\widehat{v}_{ihg}$, where $g$ is selected such that $\text{MAP}\left(\widehat{z}_{ig}\right)=1$, and $x_{ih}$ is considered good with respect to the $h$th dimension if $\widehat{v}_{ihg}>0.5$ and $x_{ih}$ is considered bad with respect to the same dimension otherwise, $i=1,\ldots,n$ and $h=1,\ldots,d$.
This is in line with the concept of snipping, complementing that of trimming, introduced in robust cluster analysis by \citet{Farc:Snip:2014} and studied in model-based clustering by \citet{Farc:Robu:2014}; for further details about snipping, refer to \citet[][Chapters~8 and 9]{Farc:Grec:Robu:2016}.
Roughly speaking, an observation is snipped when some of its dimensions are discarded but the remaining are used for clustering and estimation.





\section{Data Analyses}
\label{sec:Data Analyses}

In this section, we evaluate the performance of the proposed MSCN mixture on artificial and real data. 
Particular attention is devoted to the problem of detecting bad points.
We further provide a comparison with (unconstrained) finite mixtures of some well-established multivariate symmetric distributions. 
In detail, we compare:
\begin{enumerate}
	\item multivariate normal mixtures (MNMs);	
	\item multivariate $t$ mixtures (M$t$Ms; \citealp{Peel:McLa:Robu:2000});
	\item multivariate contaminated normal mixtures (MCNMs; \citealp{Punz:McNi:Robu:2016});
	\item multiple-scaled $t$ mixtures (MS$t$Ms; \citealp{Forb:Wrai:Anew:2014}).	
\end{enumerate}
Apart from MNMs, each mixture component of the models above has one (in the case of M$t$Ms) or more (in the case of MCNMs and MS$t$Ms) additional parameters governing the tail weight.

The whole analysis is conducted in \textsf{R} \citep{R:2018}, with all of the fitting algorithms being EM or EM variants.
MNMs are fitted via the \texttt{gpcm()} function of the \textbf{mixture} package \citep{R:mixture:2018} using the option \texttt{mnames = "VVV"}, M$t$Ms are fitted via the \texttt{teigen()} function of the \textbf{teigen} package \citep{JSS:teigen} specifying the argument \texttt{models = "UUUU"}, MCNMs are fitted via the \texttt{CNmixt()} function of the \textbf{ContaminatedMixt} package \citep{JSS:ContaminatedMixt} using the option \texttt{model = "VVV"}, while a specific \textsf{R} code
has been implemented to fit MS$t$Ms and MSCNMs.
For a fair comparison, the updates of $\bGamma_j$ and $\bLambda_j$, $j=1,\ldots,k$, for the MS$t$M are not computed with the approach discussed in \citet{Forb:Wrai:Anew:2014} but with arguments analogous to the those discussed in Section~\ref{subsec:MSCNM CM Step 2}.
To allow for a direct comparison of the competing models, all of these algorithms are initialized by providing the initial quantities $\bz_i^{(0)}$, $i=1,\ldots,n$ using the partition provided by a preliminary run of the $k$-medoids method, as implemented by the \texttt{pam()} function of the \textbf{cluster} package \citep{R:cluster:2018}.
For the competing mixture models based on the $t$ distribution, the degrees of freedom are initialized to 20.

To compare the classification results, when the true partition is available, we use the error rate (ER) and the adjusted Rand index (ARI; \citealp{hubert85}). 
The ARI corrects the Rand index \citep{rand71} for chance; its expected value under random classification is $0$, and it takes a value of $1$ when there is perfect class agreement. 

The comparison is also based on the ability of the models to detect outliers.
In this regard, the MCNMs can be used to detect outliers using an analogous procedure like the one described in Section~\ref{subsec:MSCNMs Directional outlier detection}; see \citet[][Section~5.6]{Punz:McNi:Robu:2016} for details. 
An \textit{a~posteriori} procedure (i.e., a procedure taking place once the model is fitted) to detect bad points with M$t$Ms is illustrated by \citet[][p.~232]{McLa:Peel:fini:2000}: an observation $\bx_i$ is treated as a bad point in the $j$th cluster if: 
\begin{equation}
	\sum_{j=1}^k \text{MAP}\left(\widehat{z}_{ij}\right)\delta\left(\bx_i,\widehat{\bmu}_j;\widehat{\bSigma}_j\right)
	\label{eq:MtM detection}
\end{equation} 
is sufficiently large, where $\delta(\bx_i,\widehat{\bmu}_j;\widehat{\bSigma}_j)$ is the squared Mahalanobis distance between $\bx_i$ and $\widehat{\bmu}_j$ with covariance matrix $\widehat{\bSigma}_j$, $i=1,\ldots,n$ and $j=1,\ldots,k$.
To decide how large the statistic \eqref{eq:MtM detection} must be in order for $\bx_i$ to be classified as a bad point, \citet[][p.~232]{McLa:Peel:fini:2000} compare it to the 95th percentile of the chi-squared distribution with $d$ degrees of freedom, where the chi-squared result is used to approximate the distribution of
$\delta(\bX_i,\widehat{\bmu}_j;\widehat{\bSigma}_j)$.
This procedure can be easily extended to the MS$t$M to define a strategy for the directional detection of bad points by considering the statistic:
\begin{equation}
	\sum_{j=1}^k \text{MAP}\left(\widehat{z}_{ij}\right)\delta\left(\widehat{\bGamma}^\top\left[\bx_i-\widehat{\bmu}_j\right]_h,0;\widehat{\lambda}_j\right);
	\label{eq:MStM detection}
\end{equation}
It can be compared to the 95th percentile of the chi-squared distribution with one degree of freedom in order to classify $x_{ih}$ as good or bad in the $j$th cluster, $i=1,\ldots,n$, $h=1,\ldots,d$, and $j=1,\ldots,k$.

\subsection{Synthetic Data}
\label{Subsec:artificialdata}

\begin{figure}
\centering
\includegraphics[width=0.6\textwidth]{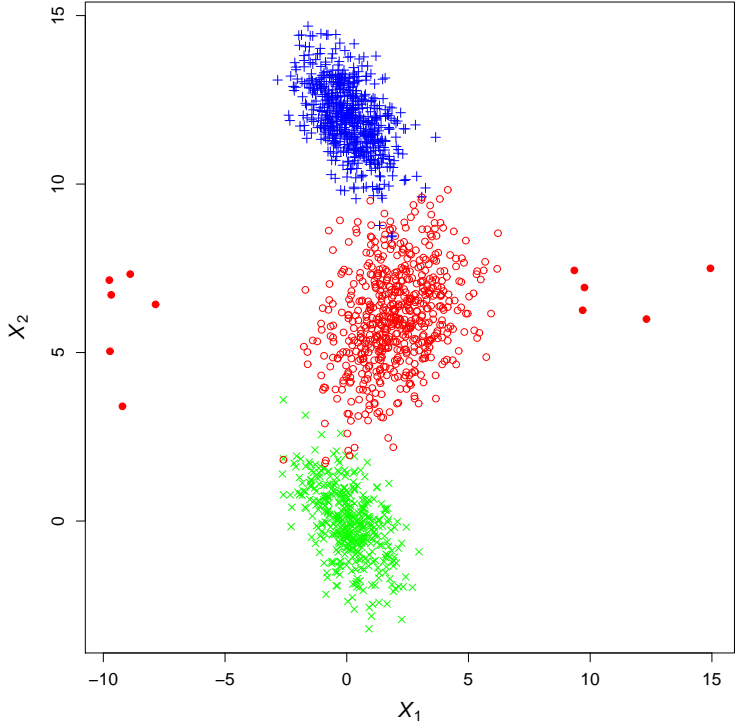}
\caption{Synthetic data: scatter plot with colors and shapes representing the different clusters.
Outliers are denoted by bullets.
\label{fig:synthetic}
}
\end{figure}

The artificial data analysis considers $n=1600$ observations, subdivided in $k=3$ groups of sizes $n_1=400$ and $n_2=n_3=600$, randomly generated by bivariate ($d=2$) normal distributions with parameters  
\begin{displaymath}
\text{$\bmu_1=\left(0,0\right)^\top$, $\bmu_2=\left(2,6\right)^\top$, $\bmu_3=\left(0,12\right)^\top$} \\
\end{displaymath}
\begin{displaymath}
\bSigma_1=\bSigma_3=
\begin{pmatrix}
1& -0.5 \\[1ex]
-0.5 &1 
\end{pmatrix},\quad \text{and} \quad \bSigma_2=
\begin{pmatrix}
2  & 0.5 \\[1ex]
0.5 & 2
\end{pmatrix}.
\end{displaymath}
Moreover, outliers have been included by substituting the first dimension ($X_1$) of 11 randomly selected points of the second cluster, with values randomly generated from a uniform distribution with support $\left(-10, -7\right) \times \left(8,15\right)$.  
\figurename~\ref{fig:synthetic} shows the scatter plot of the generated data, with colors and shapes representing the different clusters and with bullets denoting the outliers. 
As we can note, the clusters are separated sufficiently, and the outliers fall outside them; thus, we would expect the competing robust methods, directly fitted with $k=3$ components, to be able to easily recognize the underlying clusters and to detect the outliers.

\tablename~\ref{table:synthetic} shows the obtained ER and ARI values.
All of the methods have a similar classification performance.
\begin{table}[!ht]
\centering
\begin{tabular}{cccccc}
\toprule
 & {MNM} & {M$t$M} & {MS$t$M} & {MCNM} & {MSCNM}\\
\midrule 
ER &0.009 & 0.010 & 0.014 & 0.009& 0.009\\ 
  ARI &0.972 &0.97 &0.958& 0.972 &0.974\\
\bottomrule
\end{tabular}
\caption{Synthetic data: ER and ARI values for the competing mixture models with $k=3$ components}
\label{table:synthetic} 
\end{table}
\tablename~\ref{table:outsynt} reports the number of false positives (i.e. the number of points incorrectly detected as outliers) related to the outlier detection rule of the competing robust methods.
We can note how the detection rule from the MSCNM is the only one that does not incorrectly label good points as outliers.  
\begin{table}[!ht]
\centering
\begin{tabular}{cccc}
\toprule
{M$t$M}&   {MS$t$M} & {MCNM} &  {MSCNM}\\
\midrule 
 93 & 35 & 1 & 0 \\ 
 \bottomrule
\end{tabular}
\caption{Synthetic data: number of points incorrectly identified as outliers by some robust procedures}
\label{table:outsynt} 
\end{table}
On the contrary, M$t$Ms and MS$t$MS detect many more outliers than there should be. 

For the MSCNM, the estimates of the parameters $\bmu_j$ and $\bSigma_j$, $j=1,2,3$, are very close to the true ones.
Particular attention has to be devoted to evaluate the estimates of $\balpha_j$ and $\boldsymbol{\eta}_j$, $j=1,2,3$. 
Clusters 1 and 3 do not have outliers ($\widehat{\balpha}_1=\widehat{\balpha}_3=\left(0.999,0.999\right)^\top$), and cluster 2, with $\widehat{\balpha}_2=\left(0.979,0.999\right)^\top$, has about 2\% of outliers and only on the first dimension; in the same group, the bidimensional degree of contamination is $\widehat{\boldsymbol{\eta}}_2=\left(18.384,1.001\right)^\top$.
The first value of $\widehat{\boldsymbol{\eta}}_2$ highlights how the corresponding MSCN mixture component distribution needs to make its tails heavier, on the first dimension only, to accommodate the outliers included into the data.
The remaining bidimensional degrees of contamination are $\widehat{\boldsymbol{\eta}}_1=\widehat{\boldsymbol{\eta}}_3=\left(1.001,1.001\right)^\top$.
Finally, it is of interest to note that similar results are obtained for the MCNM with reference to $\bmu_j$ and $\bSigma_j$, $j=1,2,3$.
Also, in this case, the second mixture component is devoted to accommodate the outliers, with $\widehat{\alpha}_2=0.978$ and $\widehat{\eta}_2=15.096$.
However, the "omnibus" value of $\widehat{\eta}_2$ is not able to clarify that there are outliers on the first dimension only.

\subsection{Wholesale Data}
\label{subsec:Wholesale Data}

The real data analysis considers the wholesale data set, which is freely available on the UCI machine learning repository at \url{https://archive.ics.uci.edu/ml/datasets/wholesale+customers}. 
The data set originates from a larger database \citep[see][]{Abre:Anal:2011} and contains information about the annual spending, in monetary units, on $d=6$ products for $n=440$ customers of a wholesale merchant in Portugal.
The product categories are: fresh, milk, grocery, frozen, detergents paper (DP), and delicatessen.
The data set also contains two nominal variables: region (Lisboa, Porto, or other) and channel (hotel/restaurant/caf\'{e} or retail).
There is no distinguishable difference in consumption among the regions, but there is a distinguishable difference between channels.
The objective of this analysis is to segment the customers based on their spending and to compare these segments to the channel.

\begin{figure}[!ht]
\centering
\resizebox{0.9\textwidth}{!}{
\includegraphics{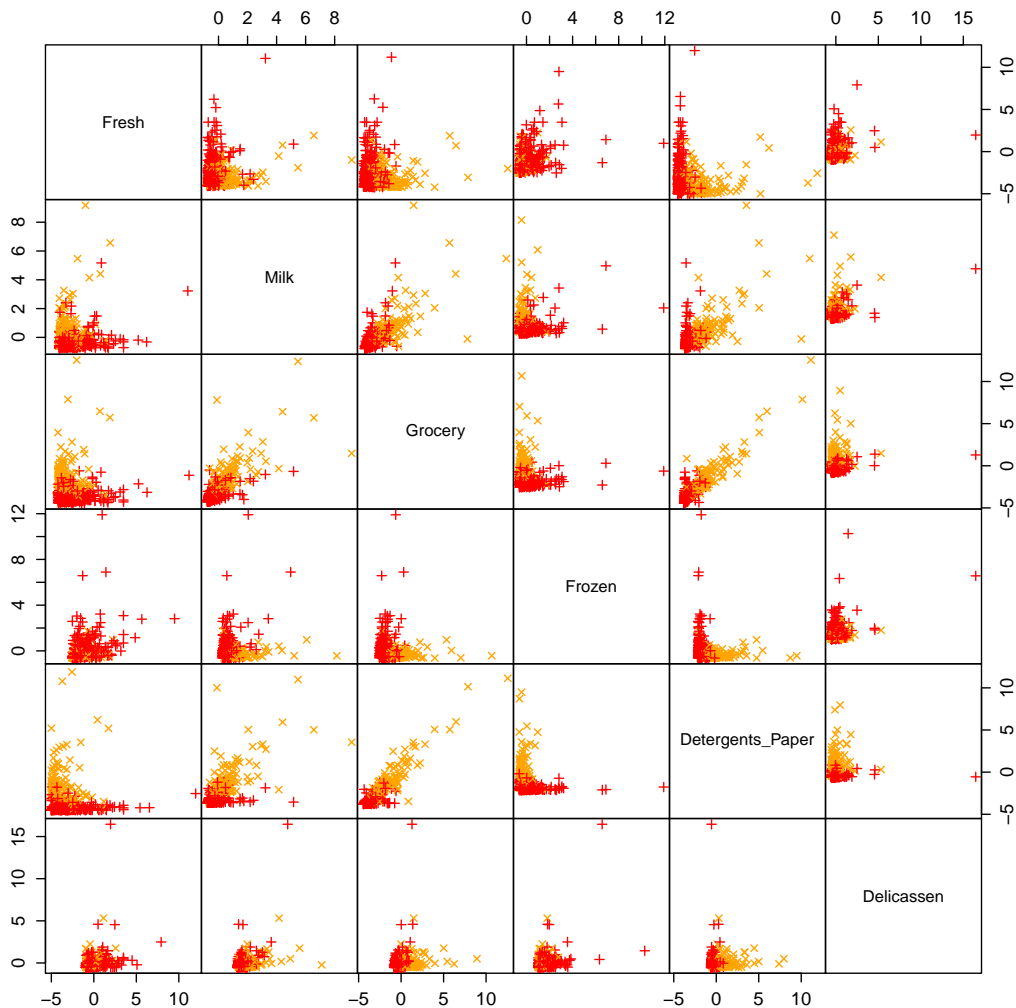} 
}
\caption{
Wholesale data: scatter plot matrix with classification of the customers based on the channel (\textcolor{orange}{$\times$} and \textcolor{red}{$+$} denote retail and horeca channels, respectively).
}
\label{fig:wholesale scatter}
\end{figure}

\figurename~\ref{fig:wholesale scatter} shows the scatter plot matrix of the standardized data, with each color and symbol representing a different channel.
There is an high level of overlap between groups, and there are different number of outliers per variable. 

The competing models were fitted with $k=2$ components, and $k$-medoids was used as the initialization strategy for all the fitting algorithms.
\tablename~\ref{table:wholesale} shows the ER for each model.
The MCNM gives the best performance with $\text{ER}=0.18$ and $\text{ARI}=0.39$.
\begin{table}[!ht]
\centering
\begin{tabular}{cccccc}
\toprule
 & {MNM} & {M$t$M}&   {MS$t$M} &{MCNM} &  {MSCNM}\\
\midrule 
ER  & 0.202 & 0.220 & 0.209 & 0.248 & 0.177 \\ 
ARI & 0.341 & 0.311 & 0.302 & 0.252 & 0.395 \\ 
\bottomrule
\end{tabular}
\caption{Wholesale data: ER and ARI values for the competing mixture models with $k=2$ components}
\label{table:wholesale} 
\end{table}
\tablename~\ref{table:outliers} shows  the number of outliers per dimension detected using the fitted MSCNM. 
Grocery, fresh, and delicatessen have the higher number of outliers.
\begin{table}[!ht]
\centering
\begin{tabular}{crrrrrrrrrrr}
\toprule
 Fresh&Milk&Grocery&Frozen&DP& Delicatessen\\
\midrule 
 5 &6 & 2 & 2& 1&4 \\
\bottomrule
\end{tabular}
\caption{Wholesale data: number of outliers per dimension detected by the MSCN mixture with $k=2$ components}
\label{table:outliers} 
\end{table}

Some of the estimated parameters from the fitted MSCNM with $k=2$ components can help in the interpretation of the results. 
\tablename~\ref{table:wholesaleparameters} shows the estimates of $\bmu_j$, $\balpha_j$, and $\boldsymbol{\eta}_j$ for each cluster.
\begin{table}[!ht]
\centering
\begin{tabular}{crrrrrrrrrrr}
\toprule
& Fresh&Milk&Grocery&Frozen&DP& Delicatessen\\
\midrule 
$\bmu_1$& 0.049 & -0.342 & -0.387 & 0.048 & -0.369 & -0.141 \\ 
$\balpha_1$ & 0.983 & 0.982 & 0.961 & 0.990 & 0.990 & 0.992 \\ 
$\boldsymbol{\eta}_1$ & 24.486 & 13.552 & 2.097 & 1.001 & 1.001 & 1.001 \\ 
\midrule 
$\bmu_2$& -0.215 & 0.935 & 1.127 & -0.228 & 1.132 & 0.366 \\ 
$\balpha_2$ & 0.987 & 0.941 & 0.971 & 0.970 & 0.947 & 0.946 \\ 
$\boldsymbol{\eta}_2$& 30.307 & 8.816 & 16.317 & 31.962 & 5.777 & 12.155 \\ 
\bottomrule
\end{tabular}
\caption{Wholesale data: some of the estimated parameters from the MSCN mixture with $k=2$ components}
\label{table:wholesaleparameters} 
\end{table} 
The customers in cluster one spend more for fresh and frozen products.
In this cluster, there is a 4\% outlying spending in grocery and 2\% in fresh and milk.
The outliers for fresh and milk are further away from the bulk of the spending for this group when compared to the outliers for grocery. 
The customers in cluster two are those spending more for milk, grocery, detergent paper, and delicatessen categories. 
There are more outliers in this cluster, and they are generally farther away from the centers when compared to cluster one, with the exception of milk.

\section{Conclusions}
\label{sec:Discussion and Future Work}

The multivariate contaminated normal (MCN) distribution, with respect to the classical multivariate normal (MN) distribution, has two additional parameters, $\alpha$ and $\eta$, denoting the proportion of good data and the degree of contamination, respectively. 
In this paper, we derived the multiple-scaled contaminated normal (MSCN) distribution to allow $\alpha$ and $\eta$ to vary across the $d$ dimensions. 
We referred to the possibility to work dimension-by-dimension using the adjective "directional."
The MSCN distribution was obtained following the strategy of \citet{Forb:Wrai:Anew:2014}.
In our setting, such a strategy was roughly based on two key elements: (1) the eigen decomposition of the scale matrix $\bSigma$ of the MCN distribution and (2) the introduction of a multidimensional Bernoulli variable indicating whether a point is good or bad separately for each dimension. 
The MSCN distribution has a closed-form representation and depends on $2d$ additional parameters, with respect to the MN distribution, which represent the proportion of good data and the degree of contamination on each dimension.
Advantageously, the MSCN distribution permits directional robust estimation of the mean vector and covariance matrix of the MN distribution and also gives  automatic directional detection of bad points in the same natural way as observations are typically assigned to the groups in the finite mixture models context, i.e., based on the posterior probabilities of being good or bad points in each dimension.
With respect to the former advantage, as an example, the estimator in \eqref{eq:AECM mu} of the mean for the generic $h$th dimension, $h=1,\ldots,d$, is a weighted mean in which the weights reduce the impact of bad points (in that dimension) in the estimation.

The MSCN distribution was applied to robust model-based clustering by introducing mixtures of MSCN distributions; a variant of the EM algorithm was also described to obtain ML estimates for the mixture parameters.
In the real and artificial data analyses of Section~\ref{sec:Data Analyses}, we demonstrated the good behavior of our directional contaminated approach when compared to mixtures of the following distributions: MN, M$t$, MCN, and MS$t$. 

Future work will focus on the following avenues:
\begin{itemize}
	\item Our mixture model implies symmetric distributions for each cluster which, under specific empirical settings, could be rather restrictive. 
This is justified by the fact that non-symmetric distributions can be approximated quite well by a mixture of several basic symmetric distributions.
While this can be very helpful for modeling purposes, it can be misleading when dealing with clustering and classification applications because one cluster may be represented by more than one mixture component simply because it has, in fact, a skewed distribution.
To overcome this issue, we could extend our MSCN distribution with the aim of introducing skewness; the resulting model could be used to define the components of a mixture.
Examples of competing approaches in this directions are given in  \citet{franczak15} and \citet{tortora18}.
	\item In the fashion of \citet{McLa:Peel:fini:2000}, \citet{McLa:Peel:Bean:Mode:2003}, and \citet{McNi:Murp:Pars:2008} for mixtures of MN distributions; \citet{McLa:Bean:BenT:Exte:2007} and \citet{Andr:McNi:Exte:2011} for mixtures of M$t$ distributions; and \citet{Punz:McNi:Robu:2014} for mixtures of MCN distributions, parsimony and dimension reduction could be obtained by exploiting local factor analyzers.
\end{itemize}

\small
\bibliographystyle{natbib}      

\end{document}